\documentclass[runningheads]{llncs}
\usepackage{graphicx}
\usepackage[font=small]{subfig}
\usepackage{multirow}
\usepackage[compatible]{algpseudocode}
\usepackage{algorithm}
\usepackage{epstopdf}
\usepackage{amsmath}

\usepackage[utf8]{inputenc}

\begin{document}

\title{Constrained Shortest Path and Hierarchical Structures\thanks{The Russian Science Foundation supports this research (Grant No. 19-71-10012. Project ``Multi-agent systems development for automatic remote control of traffic flows in congested urban road networks'')}}

\author{Adil Erzin\inst{1,2}\orcidID{0000-0002-2183-523X} \and Roman Plotnikov\inst{1,2}\orcidID{0000-0003-2038-5609} \and Ilya Ladygin\inst{3}}
\authorrunning{A. Erzin et al.}

\institute{Sobolev Institute of Mathematics, SB RAS, Novosibirsk 630090, Russia \and
St. Petersburg State University, St. Petersburg 199034, Russia  \and
Novosibirsk State University, Novosibirsk 630090, Russia\\
\email{\{adilerzin, prv\}@math.nsc.ru}, ilia.lad@mail.ru}

\maketitle              

\begin{abstract}
The Constraint Shortest Path (CSP) problem is as follows. An $n$-vertex graph is given, each edge/arc assigned two weights. Let us call them ``cost'' and ``length'' for definiteness. Finding a min-cost upper-bounded length path between a given pair of vertices is required. The problem is NP-hard even when the lengths of all edges are the same. Therefore, various approximation algorithms have been proposed in the literature for it. The constraint on path length can be accounted for by considering one edge weight equals to a linear combination of cost and length. By varying the multiplier value in a linear combination, a feasible solution delivers a minimum to the function with new weights. At the same time, Dijkstra's algorithm or its modifications are used to construct the shortest path with the current weights of the edges. However, with insufficiently large graphs, this approach may turn out to be time-consuming. In this article, we propose to look for a solution, not in the original graph but specially constructed hierarchical structures (HS). We show that the shortest path in the HS is constructed with $O(m)$-time complexity, where $m$ is the number of edges/arcs of the graph, and the approximate solution in the case of integer costs and lengths of the edges is found with $O(m\log n)$-time complexity. The a priori estimate of the algorithm's accuracy turned out to depend on the parameters of the problem and can be significant. Therefore, to evaluate the algorithm's effectiveness, we conducted a numerical experiment on the graphs of roads of megalopolis and randomly constructed unit-disk graphs (UDGs). The numerical experiment results show that in the HS, a solution close to optimal one is built 10--100 times faster than in the methods which use Dijkstra's algorithm to build a min-weight path in the original graph.

\end{abstract}

\section{Introduction}
In the modern communication networks, it is not enough to find an optimal path using one characteristic of its elements. To meet Quality of Service (QoS) requirements, it is often necessary to take into account more than two characteristics at the same time (cost, length, delay, reliability, etc.) \cite{Koster:14}. We set the following problem. Given a weighted directed graph $G=(V,A)$, where $V$ is the set of vertices ($|V|=n$), $A$ is the set of arcs ($|A|=m$), and each arc $a\in A$ has length $l(a)$ and cost $c(a)$. It is necessary to find a min-cost path between a given pair of vertices $s$ and $t$ ($s-t$ path) of length at most $\beta$. In the literature, this problem is called the Constrained Shortest Path (CSP) problem. CSP is NP-hard, both in the general case \cite{Garey:79} and in the case of acyclic networks \cite{Wang:96}. Both exact exponential \cite{Kuipers:02,Lozano:13,Widyono:94} and approximation polynomial algorithms \cite{Hassin:92,Juttner:01,Lorenz:00,Orda:99,Wang:16,Wang:96,Xiao:05} are proposed to solve it.	

The exact Constrained Bellman-Ford (CBF) algorithm proposed in \cite{Widyono:94} has exponential complexity, but it is faster than brute force on average. The main idea behind this algorithm is to systematically search for the least cost paths while monotonically increasing length. First, the algorithm finds a min-cost $s-t$ path. Next, for each vertex $u$, a list of min-length paths from $s$ to $u$ is created. Then, a vertex is selected that lies on the $s-t$ path with minimum cost, the list of which contains the path that satisfies the constraint. The algorithm then explores the neighbors of this vertex using breadth-first search \cite{Cormen:00}, and (if necessary) adds new paths to the lists of neighbors.  This process continues as long as the length constraint is met and there is a path for further exploration.

Another exact algorithm is the Pulse algorithm proposed in \cite{Lozano:13}. Its essence is to apply an impulse from vertex $s$ to the neighboring vertices, then from all neighboring to the next neighbors, etc. Each time, the following characteristics of the partial path are stored in memory: the vertices passed, the value of the objective function, and the current length. When the impulse reaches vertex $t$, then the constructed path along with all the characteristics is stored. In this way, all possible paths can be found, including optimal. The difference between this algorithm and the full enumeration lies in the special strategies for cutting off partial paths ("pulses"). In the paper, these strategies are \emph{dominance}, \emph{bounds} and \emph{infeasibility}. The essence of the dominance strategy is to remember the best paths in terms of cost and length, bounds strategies — in the systematic pruning of paths with a worse objective function than the paths already found. Infeasibility allows to cut off pulses that are unpromising in length at an early stage (this is achieved by calculating the shortest distance from $t$ to each other vertex).

Hassin in \cite{Hassin:92} proposed two $\varepsilon$-approximation algorithms for the case of positive arc weights with costs $O((\frac{mn}{\varepsilon} + 1)\log\log B)$ and $ O(\frac{mn^2}{\varepsilon}\log\frac{n}{\varepsilon})$, where $B$ is an upper bound on the path cost. The first algorithm uses upper and lower bounds ($UB$ and $LB$ respectively). At the start of the program, they are given the values $LB=1$, $UP$ is the sum of $(n-1)$ largest arc costs. Then, using a special testing procedure, the estimates are systematically improved, and, using the results obtained, new arc costs are set in the form $c^\prime (u,v)=\lfloor\frac{c(u,v)(n-1)}{\varepsilon LB}\rfloor$ $\forall (u,v)\in E$, which allows us to obtain the required path. Orda \cite{Orda:99} and Lorenz et al. \cite{Lorenz:00} modified $\varepsilon$-approximation algorithms to scale better in hierarchical networks.

A special place among the approximation algorithms is occupied by backward-forward heuristic (BFH). First, for each vertex $u\in V$, two $u-t$ paths are searched — min-cost path and min-length path. This can be done, for example, with the Reverse-Dijkstra \cite{Ahuja:93} algorithm. Then, starting from the vertex $s$, a modification of Dijkstra's algorithm is applied, in which an additional condition is used to relax the arc, using the previously found paths (arc relaxation has the same meaning as in the usual Dijkstra algorithm). Examples of algorithms using similar approaches have been given by Reeves and Salama \cite{Reeves:00}, Sun and Langendorfer \cite{Sun:98}. A similar algorithm for a multi-constrained problem was proposed by Ishida \cite{Ishida:98}.

Especially for large road networks, Wang et al. \cite{Wang:16} developed the constrained labeling algorithm COLA. It is based on two special properties that are characteristic for large road networks. First, road networks are usually (roughly) planar, which makes it possible to effectively divide the graph into several subgraphs with special boundary vertices, between which it is required to find a path inside each subgraph. Secondly, often in the solutions of CSP problems on road networks there are a small number of landmarks \cite{Goldberg:05} — vertices that are present in valid paths much more often than others. According to experiments, an algorithm that takes into account these properties copes with mainland-sized road networks many times better than other algorithms.

In the generalization of the CSP problem -- Multi-Constrained Path (MCP) problem -- each arc $a\in A$ has $p$ parameters and it is required to find a path that satisfies $p$ constraints with respect to each of the parameters. A summary and comparison of algorithms that solve MCP can be found in \cite{Kuipers:02}.

\subsection{Our contribution}
Our approach to solving the CSP problem is based on the Lagrange Relaxation Aggregated Cost (LARAC) algorithm developed in \cite{Juttner:01} and summarized in \cite{Xiao:05}. In this approach, the Lagrange multiplier $\alpha >0$ is introduced, and instead of the cost $a_{ij}$ and the length $b_{ij}$ of the arc $(i,j)$, one aggregated weight $c_{ij}=a_{ ij}+\alpha b_{ij}$ is used. For a fixed value of $\alpha$, a path $P(\alpha)$ of minimum weight $c(\alpha)$ is constructed, which has cost $a(\alpha)$ and length $b(\alpha)$. If the length of the path exceeds the allowable value, then the value of $\alpha$ increases. Otherwise, it decreases.

To reduce the complexity, we propose to build special hierarchical structures (HS) according to a given graph, in which copies of the same vertex can be located at several neighboring levels, and arcs connect the vertices of only neighboring levels. However, in the HS, the sink $t$ is incident to the arcs from all adjacent vertices, regardless of the level of their location. Further, using heuristic considerations, additional arcs are added to the HS instead of some paths in the original graph.

We have shown that the shortest path in the HS is constructed with $O(m)$-time complexity, where $m$ is the number of arcs/edges in the original graph. If the graph is sparse, then this is a big gain compared to the Dijksra's algorithm and its modifications. Obviously, not all arcs of the original graph are included in the HS, so the found path may differ from the shortest one. To compare the running time and the accuracy of our approach, a numerical experiment was carried out. The simulation shows that the construction of the shortest path in HS is several times faster than if one use traditional algorithms. At the same time, in the HS, solutions close to optimal are constructed.

\vspace{1cm}
The rest of the paper is organized as follows. In the next section, we present the mathematical formulation of the CSP problem. In the third section, we present the rules for constructing different hierarchical structures. Section 4 is devoted to the description of algorithm $A_\alpha$, which ideologically coincides with the LARAC \cite{Juttner:01,Xiao:05} and builds an approximate solution to the problem. This section also provides estimates for the running time and accuracy of the $A_\alpha$. Section 5 describes the numerical experiment, as well as the results of simulation. The last section concludes the paper.

\section{Problem formulation}
Let a mixed graph $G=(V,A)$, $|V|=n$, $|A|=m$, be given, whose arcs/edges we will call the \emph{arcs} for definiteness. Each arc $(i,j)\in A$ is assigned two non-negative numbers: \emph{cost} $a_{ij}$ and \emph{length} $b_{ij}$. We assume that the graph does not contain a pair of vertices $i,j\in V$ connected by a simple path $P_{ij}$ in which all internal vertices (that is, not coinciding with $i$ and $j$) have degree 2. If such a path exists, then instead of it we add one arc $(i,j)$, the cost of which is equal to the sum of the costs $a_{ij}=\sum\limits_{(p,q)\in P_{ij}} a_{pq}$, and the length is equal to the sum of the lengths $b_{ij}=\sum\limits_{(p,q)\in P_{ij}} b_{pq}$ of the arcs included in it. It is required to find a path from vertex $s\in V$ to vertex $t\in V$ ($s-t$ path) of minimum cost and length no more than $\beta >0$. If $\Pi$ is a set of simple $s-t$ paths, then it is required to find the path $P\in\Pi$, which is the solution to the following problem.
\begin{equation}\label{e1}
  \sum\limits_{(i,j)\in P}a_{ij}\rightarrow\min\limits_{P\in\Pi};
\end{equation}
\begin{equation}\label{e2}
  \sum\limits_{(i,j)\in P}b_{ij}\leq\beta.
\end{equation}

As noted above, the problem (\ref{e1}) is polynomially solvable, and the problem (\ref{e1})-(\ref{e2}) is NP-hard even if the arc lengths are the same \cite{Garey:79}.

\section{Hierarchical structures}
\begin{figure}
\centering
\includegraphics[bb= 0 70 620 500, clip, scale=0.5]{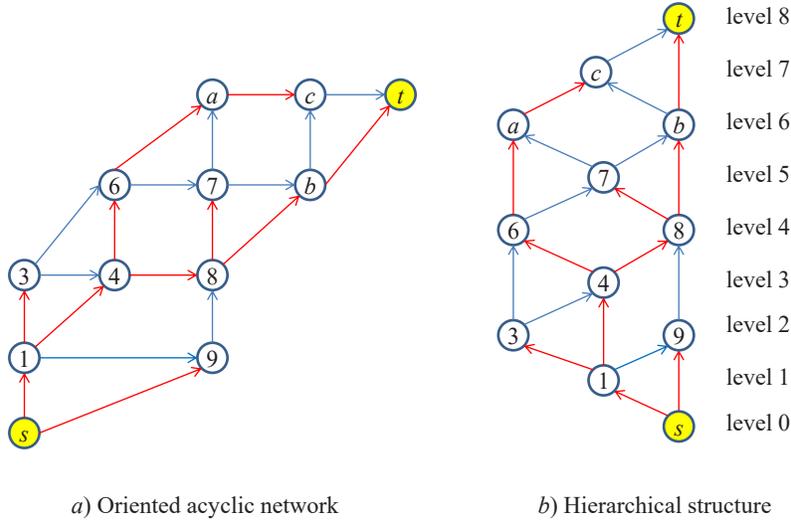}
\caption{HS for directed acyclic graph}
\label{fig1}
\end{figure}

First, consider a directed acyclic graph (Fig. \ref{fig1}\emph{a}) with one non-negative weight assigned to each arc.
If the vertices $s$ and $t$ are known, then the HS is constructed in this case without loss of arcs as follows. On level 0 we place the vertex $s$. Next, on the level $l\geq 1$ we place vertices where there is a path consisting of $l$ arcs, but there is no path with $l+1$ or more arcs (Fig. \ref{fig1}\emph{b} ). As a result, the destination vertex $t$ gets to some last level $L$. If there was an arc $(i,j)$ in the original graph $G$, then the same arc exists in the HS. In this case, it is obvious that the vertex $j$ is on a level with a higher number compared to the level number of the location of the vertex $i$. Moreover, in the process of building a HS, we can simultaneously build the shortest paths to each vertex (see the red arcs in Fig. \ref{fig1}). To do this, we consider in turn the vertices of the levels $1,\ldots,L$. Among the arcs entering the vertex $i$, which is at the level $l$, choose one that belongs to the shortest path going from $s$ to $i$. This is easy to implement by storing the length of the shortest path to every vertex adjacent to $i$ that is on a level less than $l$. As a result, the shortest $s-t$ path will be constructed with $O(m)$-time complexity.

\begin{figure}
\centering
\includegraphics[bb= 0 30 700 500, clip, scale=0.5]{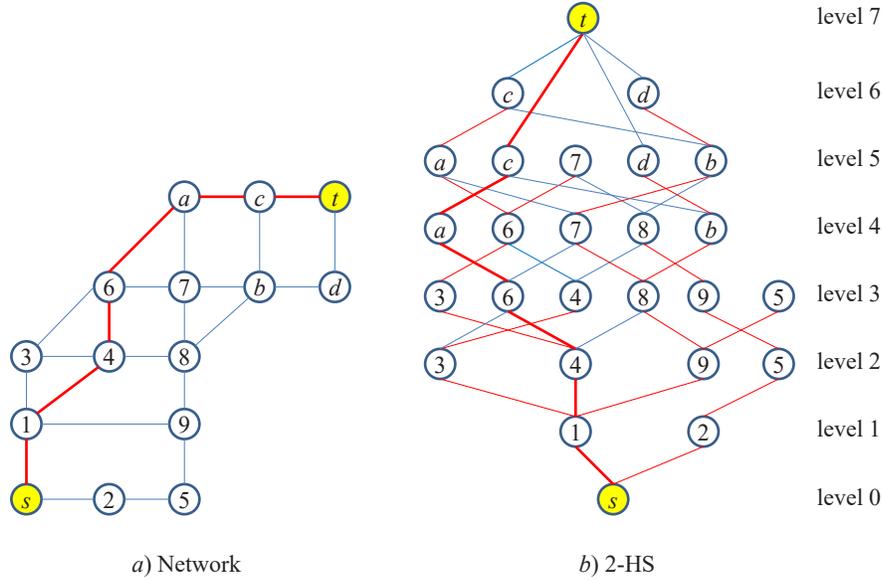}
\caption{2-HS for the mixed graph}
\label{fig2}
\end{figure}
If graph $G$ is arbitrary (not directed acyclic), then the placement of vertices at the levels of the HS is ambiguous. In this case, we will build $k$-HS, where $k$ is a positive integer constant. In the $k$-HS, $k$ copies of each vertex $i$ are located at the levels $l,l+1,\ldots,l+k-1$, where $l$ is the minimum number of edges in the path from $s$ to $i$ (see example of 2-HS in Fig. \ref{fig2}). In the $k$-HS, the arcs link only the vertices of neighboring levels, except for the vertex $t$, which is connected with all adjacent vertices, regardless of their placement level (Fig. \ref{fig2}\emph{b}).

In the example in Fig. \ref{fig2}\emph{b} the images of vertex 4 are located at level 2 and at level 3. Moreover, the arc $(1,4)$ enters vertex 4 at level 2, and the arc $(3,4)$ enters vertex 4 at level 3. Some vertices in HS may turn out to be dead ends -- no arcs go out of them. On Fig. \ref{fig2}\emph{b} such vertices are 5 at level 3 and 7 at level 5.

If it is required to build paths of minimum weight to all HS vertices, then this can be done during the construction of $k$-HS, similarly to the method described above. For this, the vertices of levels $1,\ldots,L$ are considered in turn. For an arbitrary vertex $i$ of the $l$-th level, an incoming arc $(p,i)$ is chosen such that $c_{pi}+d_p=\min\limits_q (c_{qi}+d_q)$, where the vertex $q$ is at level $l-1$, and $d_q$ is the minimum weight of the path from $s$ to $q$ (it was found earlier when looking at the vertices of level $l-1$). In the example in Fig. \ref{fig2}\emph{b} arcs included in shortest paths are red. Since $k=const$, the time complexity of finding such paths is still $O(m)$.

In 1-HS, each vertex goes to one level, the number of which is equal to the minimum number of arcs in the path from $s$. As a result, all $s-t$ paths in the 1-HS consist of the minimum number of arcs. In the $k$-HS, $k>1$, the number of arcs in the $s-t$ paths, as well as the paths themselves, is greater, which makes it possible to find a path better than in the 1-HS. On Fig. \ref{fig2} the best path (for example, the shortest one) is indicated by bold red edges.

Since not all promising paths fall into the $k$-HS, then when the nodes are the points in the plane, we add some arcs to the $k$-HS that connect vertices of non-adjacent levels based on the following heuristic. For each vertex $v\in V$, we choose a \emph{perspective} arc $a(v)$ outgoing from $v$ in the direction of the sink $t$, defined by equation $a(v)=\arg\max_{( ij)\in A}{|\vec{ij}|\cos(\angle(\vec{ij}, \vec{it})/c_{ij}}$ if it is greater than 0. For each vertex $v\in V$ and any value $p\in [1,n-1]$ a path (if it exists) outgoing from $v$ and consisting of $p$ perspective arcs can be uniquely defined. For a fixed $p_{max}$, we connect the vertices in the $k$-HS that are the ends of the perspective paths of length $p=2,\ldots,p_{max}$. Thus, in the $k$-HS$p_{max}$ no more than $n p_{max}$ additional arcs are added.

\section{Algorithm $A_\alpha$}
The algorithm presented below essentially coincides with LARAC \cite{Juttner:01,Xiao:05}, but we will describe it in the following interpretation convenient for us. Instead of two characteristics of each arc $(i,j)\in A$: cost $a_{ij}$ and length $b_{ij}$, we introduce one aggregated characteristic equal to $c_{ij}(\alpha)=a_{ij }+\alpha b_{ij}$, $\alpha\geq 0$, which we call the \emph{weight} of the arc. Denote by $P(\alpha)$ min-weight $s-t$ path when the weights of the arcs are equal to $c_{ij}(\alpha)$, $(i,j)\in A$. Its cost is $a(\alpha)$ and its length is $b(\alpha)$. If the path $P(0)$ is feasible, i.e. the inequality (\ref{e2}) $b(0)\leq\beta$ is satisfied, then this is the optimal path. Otherwise, the value of $\alpha$ should be increased until we find the minimum $\alpha=\alpha^*$ for which the constructed $s-t$ path $P(\alpha^*)$ is feasible. To find $\alpha^*$ one can apply a dichotomy algorithm. The authors in \cite{Juttner:01,Xiao:05} use an alternative way to change $\alpha$ values.

\begin{figure}
\centering
\includegraphics[bb= 0 30 700 300, clip, scale=0.5]{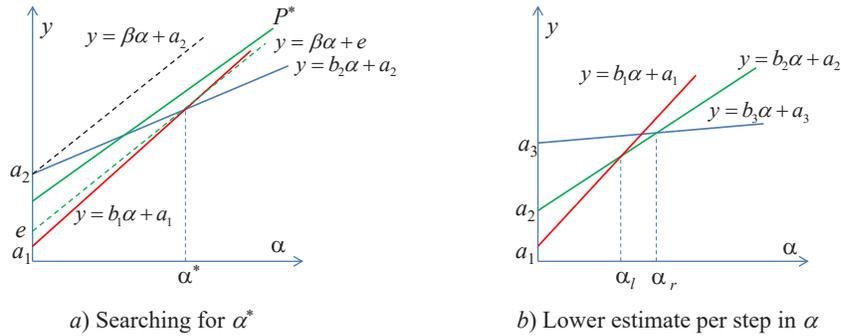}
\caption{Representation of $s-t$ paths as straight lines on the plane $(\alpha,0,y)$}
\label{fig3}
\end{figure}
Of course, $P(\alpha^*)$ will not always be the optimal solution to the problem (\ref{e1})-(\ref{e2}) (see Fig. \ref{fig3}\emph{a}). However, in the case of integer costs and arc lengths, one can estimate the accuracy of the resulting solution, as well as the number of steps to search for $\alpha^*$. Indeed, an arbitrary $s-t$ path on a plane with a horizontal coordinate axis $\alpha$ and a vertical axis $y$ is characterized by a straight line of the form $y=b\alpha+a$. The length of the path $b$ determines the slope of the straight line, and the cost of the path determines the point of intersection of the straight line with the vertical axis. The entire family of $s-t$ paths forms a minorant of straight lines whose slope angle decreases with increasing $\alpha$. It is required to find the minimal $\alpha=\alpha^*$ for which the minorant is determined by the straight line $y=b\alpha^*+a$ with $b\leq\beta$. The path $P(\alpha^*)$ corresponding to this line will be an approximate solution to the problem (\ref{e1})-(\ref{e2}). Let us assume that $\alpha^*$ is determined by the intersection of the lines $y=b_1\alpha+a_1$ and $y=b_2\alpha+a_2$, $b_2<\beta$, $b_1>\beta$. From the equality $b_1\alpha^*+a_1=b_2\alpha^*+a_2$, we get that $\alpha^*=\frac{a_2-a_1}{b_1-b_2}$. If $a_{ij}\leq A$ and $b_{ij}\leq B$, then $\alpha^*\leq An$.

Assume that the line corresponding to the optimal path passes above the minorant at the point $\alpha^*$ (green line in Fig. \ref{fig3}\emph{a}). Then the parameter $e$ of the line $y=\beta\alpha+e$ passing through the point of intersection of the lines $y=b_1\alpha+a_1$ and $y=b_2\alpha+a_2$ (dashed green line in Fig. \ref{fig3}\emph{a}), is the lower bound for the optimum. We have $\beta\alpha^*+e=b_2\alpha^*+a_2$, whence $e=a_2-(\beta-b_2)\alpha^*$. Therefore, taking into account the integer parameters, the ratio
$$
 \varepsilon\leq\frac{a_2}{e}=\frac{a_2}{a_2-(\beta-b_2)\frac{a_2-a_1}{b_1-b_2}}\leq\frac{1}{1-\frac{\beta-a_2}{b_1-b_2}}
 \leq\frac{1}{1-\frac{\beta-1}{\beta}}\leq\beta.
$$

As mentioned above, the dichotomy method can be used to find $\alpha^*$. Let us estimate the number of iterations of the method for integer parameters of the problem. An upper estimate for the value of $\alpha^*$ was obtained above. Let us find a lower estimate per step in $\alpha$ when all parameters of the problem are integers. To do this, take three lines $y=b_1\alpha+a_1$, $y=b_2\alpha+a_2$, $y=b_3\alpha+a_3$, $b_1>b_2>b_3$, $a_1<a_2<a_3$, which form two neighboring break points $\alpha_l$ and $\alpha_r$, $\alpha_l<\alpha_r$, of the minorant. We have $b_1\alpha_l+a_1=b_2\alpha_l+a_2$ and $b_3\alpha_r+a_3=b_2\alpha_r+a_2$. Consequently,
$$
 \alpha_r-\alpha_l=\frac{a_3-a_2}{b_2-b_3}-\frac{a_2-a_1}{b_1-b_2}\geq\frac{1}{(b_2-b_3)(b_1-b_2)}>
 \frac{1}{b_1b_2}\geq\frac{1}{B^2n^2}.
$$
If $K$ denotes the maximum number of iterations in the dichotomy method, then $An/2^K\leq 1/B^2n^2$. Hence $K=O(\log n)$.

If, for each value of $\alpha$, Dijkstra's algorithm is used to find the min-weight path, then the complexity of constructing an approximate solution to the problem (\ref{e1})-(\ref{e2}) is $O(n^2\log n)$ . If we look for the min-weight path in the HS, then the complexity of obtaining an approximate solution is $O(m\log n)$.

The resulting guaranteed accuracy is rough. Therefore, we conducted a numerical experiment in which we compared both the running time of the algorithm and the accuracy of the solution being constructed. The results of the numerical experiment are presented in the next section.

\section{Simulation}
We implemented the proposed algorithm using the programming language C++. We also constructed the ILP model described in \cite{Handler:80} and use GUROBI software for its solving. As test instances we used the road map of New York taken from \cite{9dimacs} and randomly generated unit disk graphs (UDG). The experiment was carried out on the AMD Ryzen 5 3550H 2.1 GHz 8 Gb RAM, Windows 10x64.

There are two weights defined for each arc in the data set of the New York road map. The first weight is the distance between nodes, and the second weight is the average traveling time. To avoid large values we divided all parameters by 100 and left only the integer parts.

UDGs were constructed in a following way. At first, a set of nodes were randomly uniformly spread over a squared region. Then, each two nodes were connected by two oppositely directed arcs iff the distance between them is less than predefined value $r$. After that, two weights were defined for each arc. The first weight equals to the distance, and the second weight equals to the distance multiplied by noise factor -- a random real value uniquely generated for each arc and uniformly distributed in the interval $[1,3]$.

Practically, actual running time spent to find one-weight shortest path (SP) depends on the proximity between source and target. That is why we consider separately instances when the distance between source and target is small (25 $\%$ of graph diameter), medium (50$\%$), and large (75$\%$). For each graph instance and each variant of distance between source and target, we generated 10 random problem instances.

We tested different variants of HS based heuristic in order to find better combination of its parameters. To be precise, for each $k=1,2,3$ and $p_{max}=1,2,3$ we run $k$-HS$p_{max}$ on each test instance and compared their performance with Dijkstra's algorithm (to be short, it is called Dij below). Also, we used each heuristic that solves SP problem in the LARAC based algorithm that approximately solves CSP. Note that for the LARAC based approach we used the rules of updating $\alpha$ from \cite{Juttner:01}. To denote these algorithms the prefix A\_ is used.

In Fig. \ref{fig-expres12} the results on the New York map are presented. Here and in the next figures, the average values among launching the algorithms on 10 random instances are presented, and the vertical intervals stand for the standard deviations. On the one hand, as it is seen in Fig. \ref{fig-expres1}, the HS based algorithms bring significant performance error, but, on the other hand, it noticeably decreases with growth of $k$ and $p_{max}$, and, according to Fig. \ref{fig-expres2}, these heuristics spend less time than Dijkstra's algorithm. Of course, time cost also increases with growth of $k$ and $p_{max}$, so the moderate values of these parameters may be chosen to achieve less quality degradation with significant speedup.

\begin{figure}[h]
\centering
\subfloat[\label{fig-expres1}Ratio]{\includegraphics[width=0.45\textwidth]{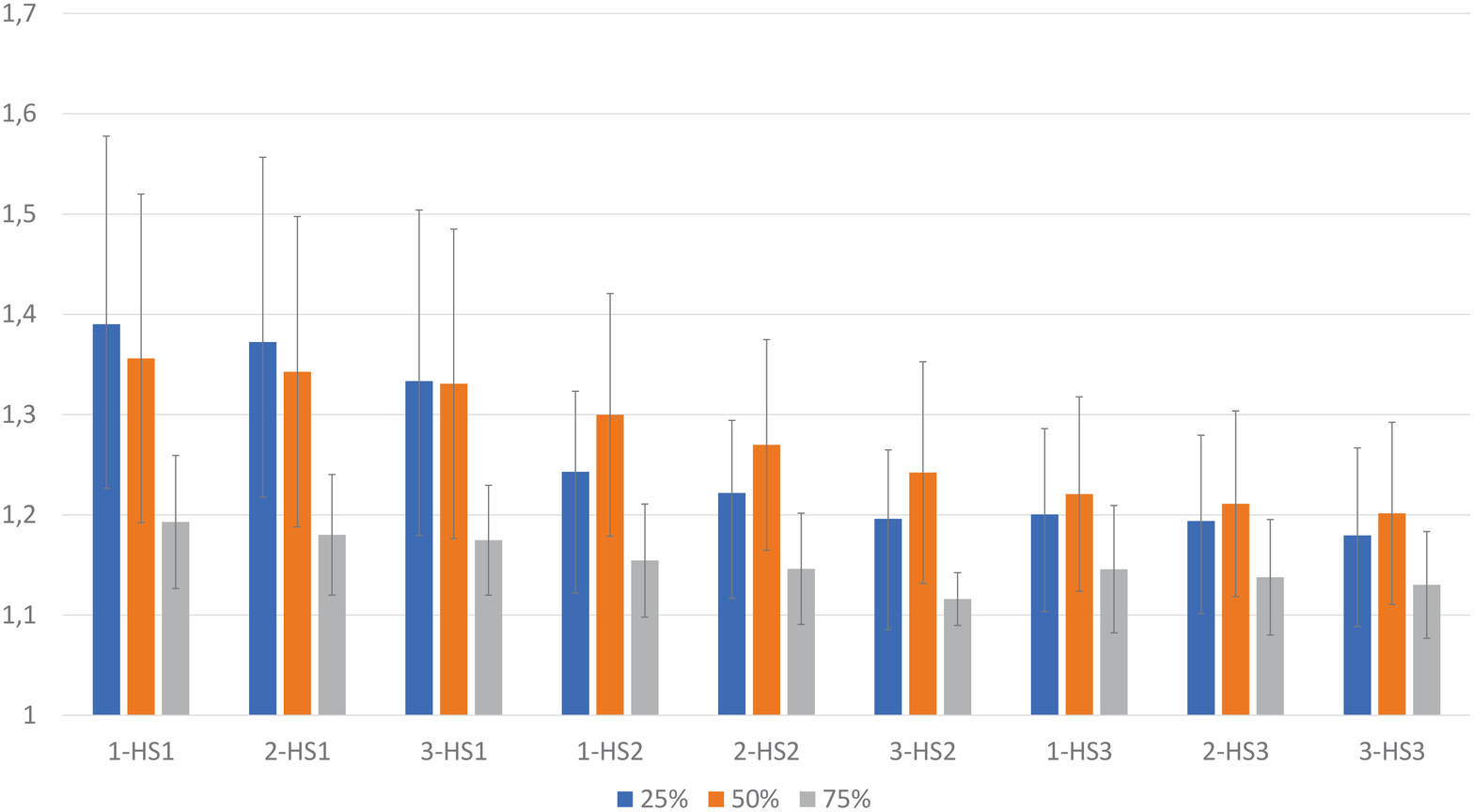}} \hfill
\subfloat[\label{fig-expres2}Time in seconds]{\includegraphics[width=0.45\textwidth]{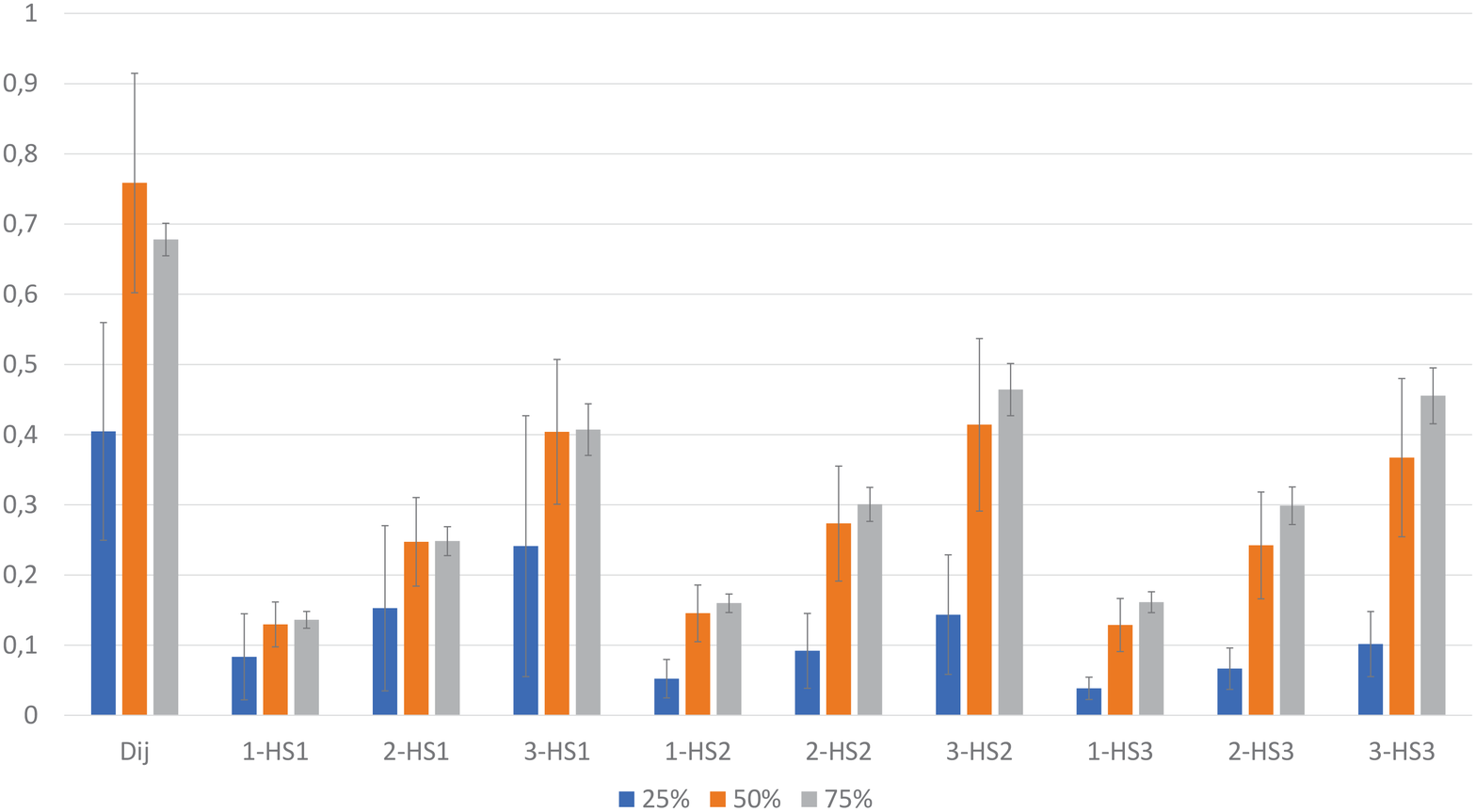}} \hfill
\caption{Algorithms results for the SP problem on the New York map. Average values and standard deviations.} \label{fig-expres12}
\end{figure}

\begin{figure}[h]
\centering
\subfloat[\label{fig-expres3}Path length]{\includegraphics[width=0.45\textwidth]{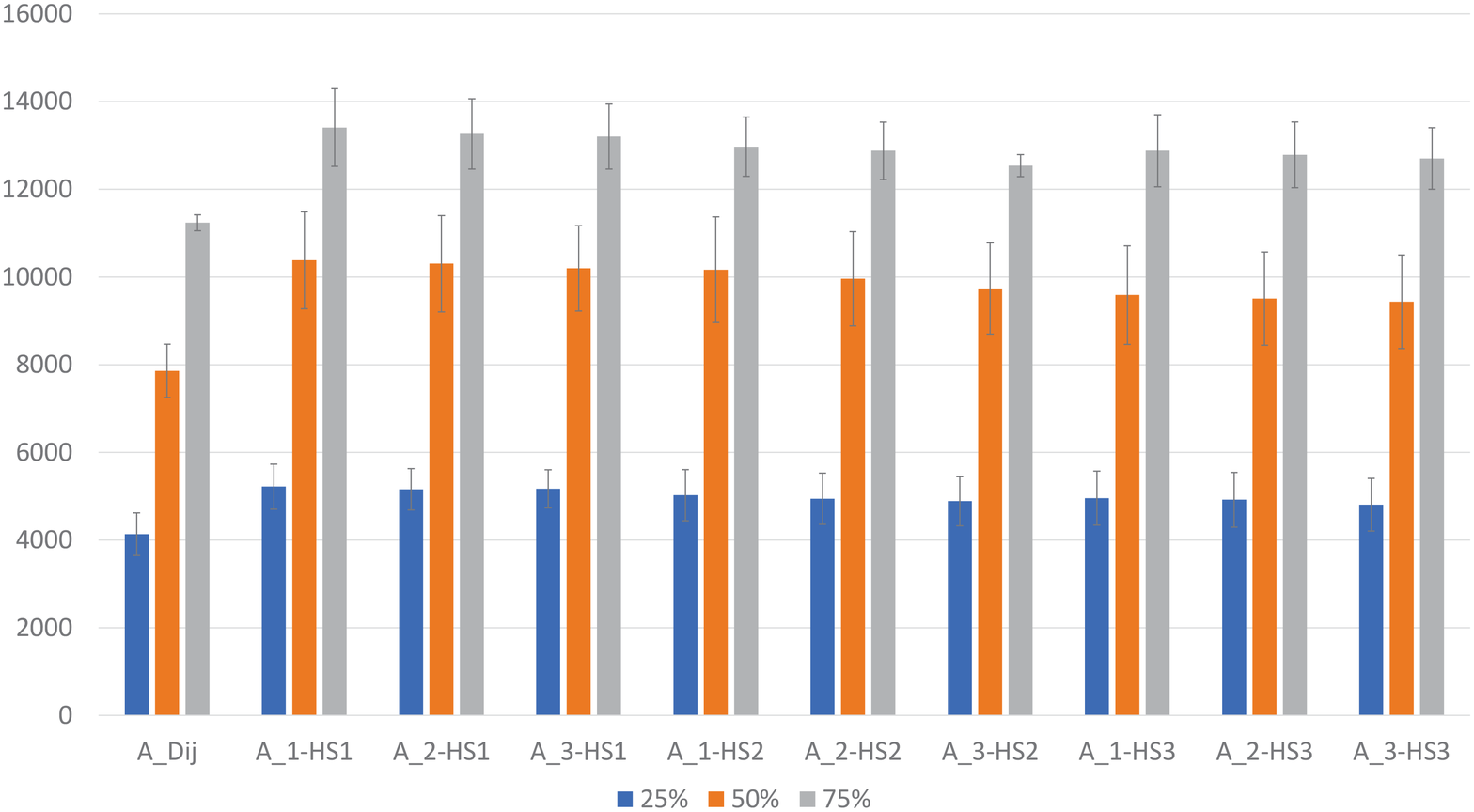}} \hfill
\subfloat[\label{fig-expres11}Ratio]{\includegraphics[width=0.45\textwidth]{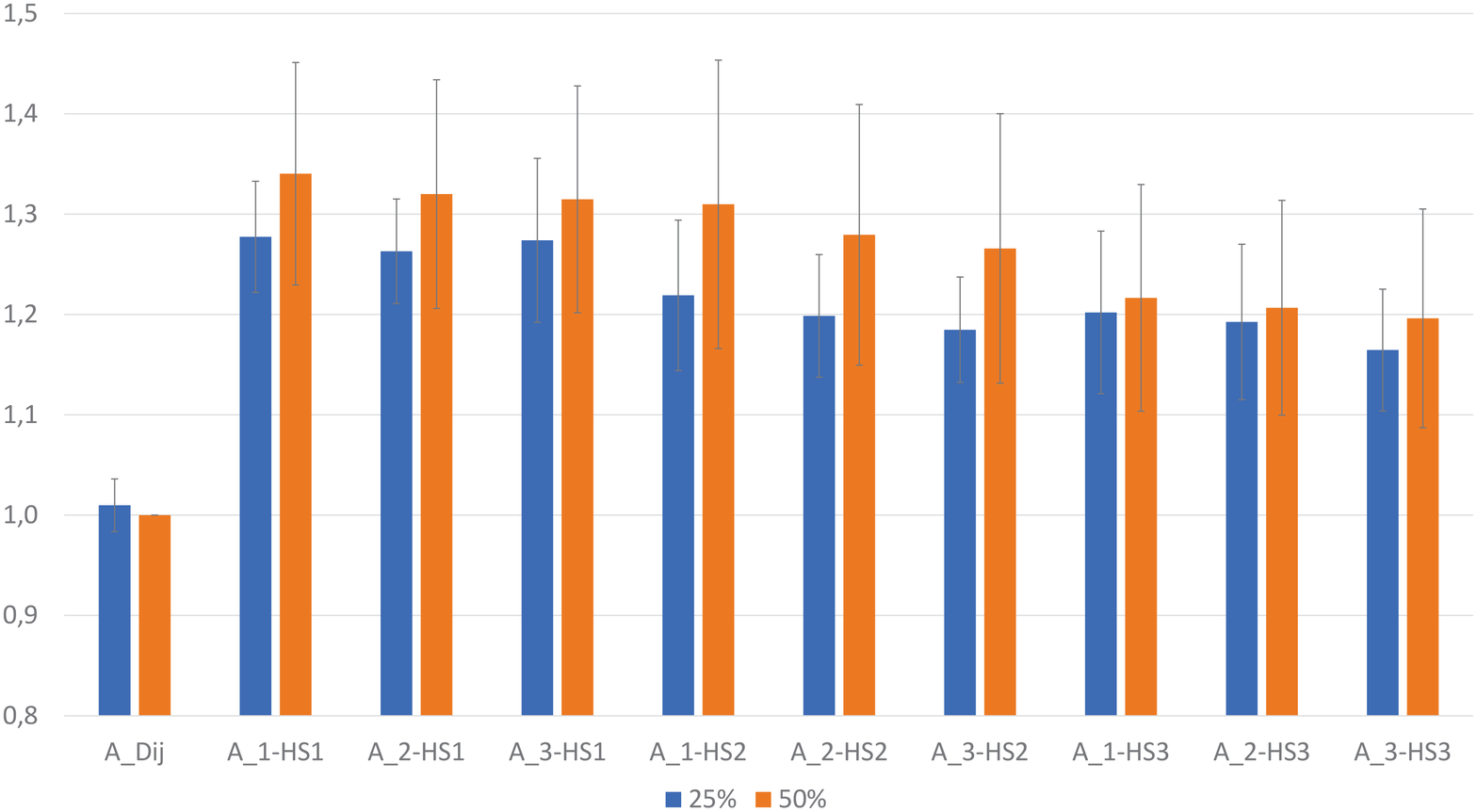}} \hfill
\subfloat[\label{fig-expres4}Time in seconds]{\includegraphics[width=0.45\textwidth]{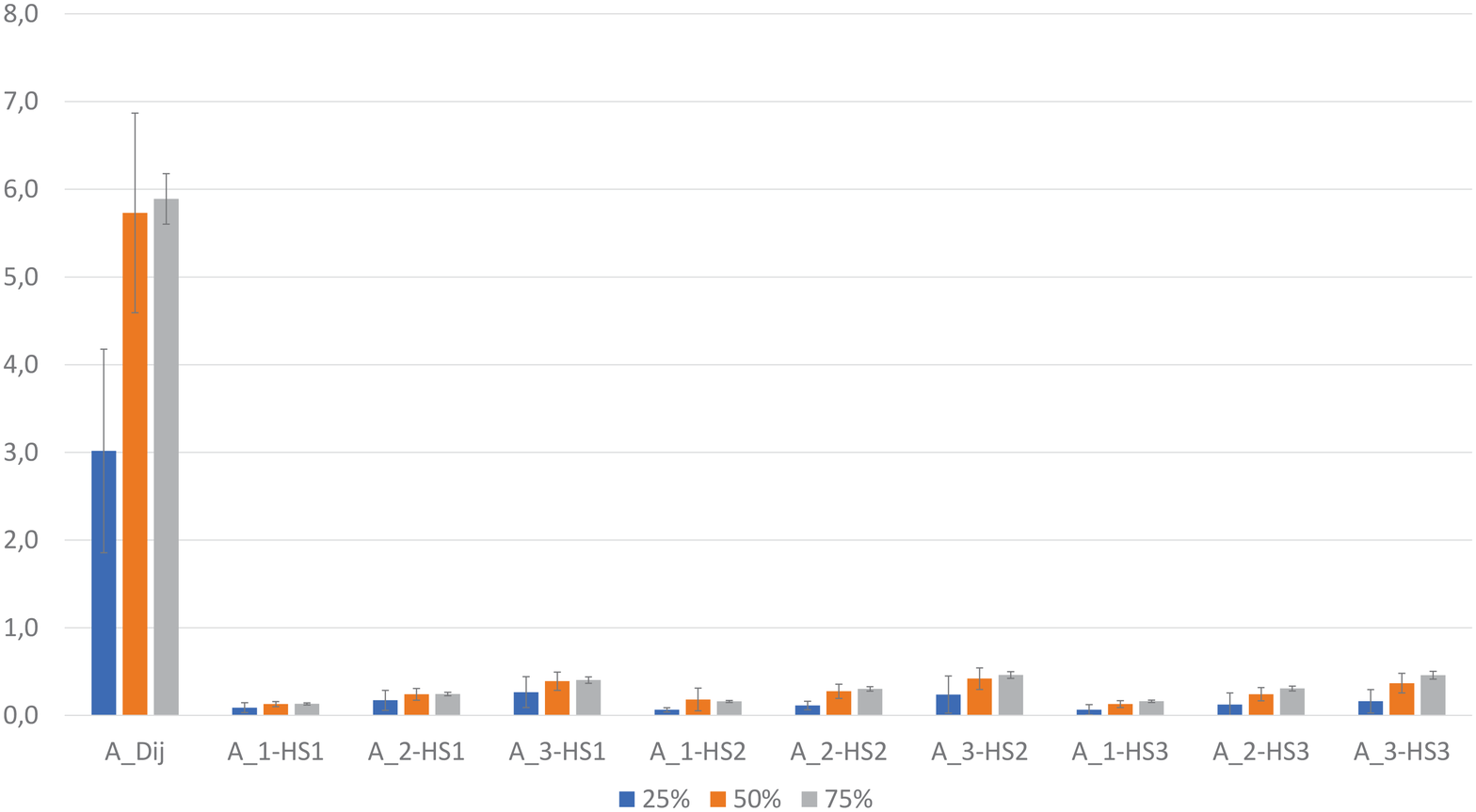}} \hfill
\caption{Algorithms results for the CSP problem on the New York map. Average values and standard deviations.} \label{fig-expres34}
\end{figure}

The results of application of these algorithms to the LARAC based approach for CSP are presented in Fig. \ref{fig-expres34}. The average path lengths are presented in Fig. \ref{fig-expres3}, the average ratio values that was obtained on the cases when GUROBI found optimal solution are presented in Fig. \ref{fig-expres11}, and Fig. \ref{fig-expres4} shows the average running time. Note that GUROBI failed to find solution to the large-size instances, when the distance between $s$ and $t$ is 75$\%$ of the graph metric diameter. Here, again, one can observe that using Dijkstra's algorithm allows to get more precise solution on average but $HS$ based heuristics allow to find approximate solution 10--100 times faster.

\begin{figure}[h]
\centering
\subfloat[\label{fig-expres5}$n = 10000$, $r = 0.1$]{\includegraphics[width=0.45\textwidth]{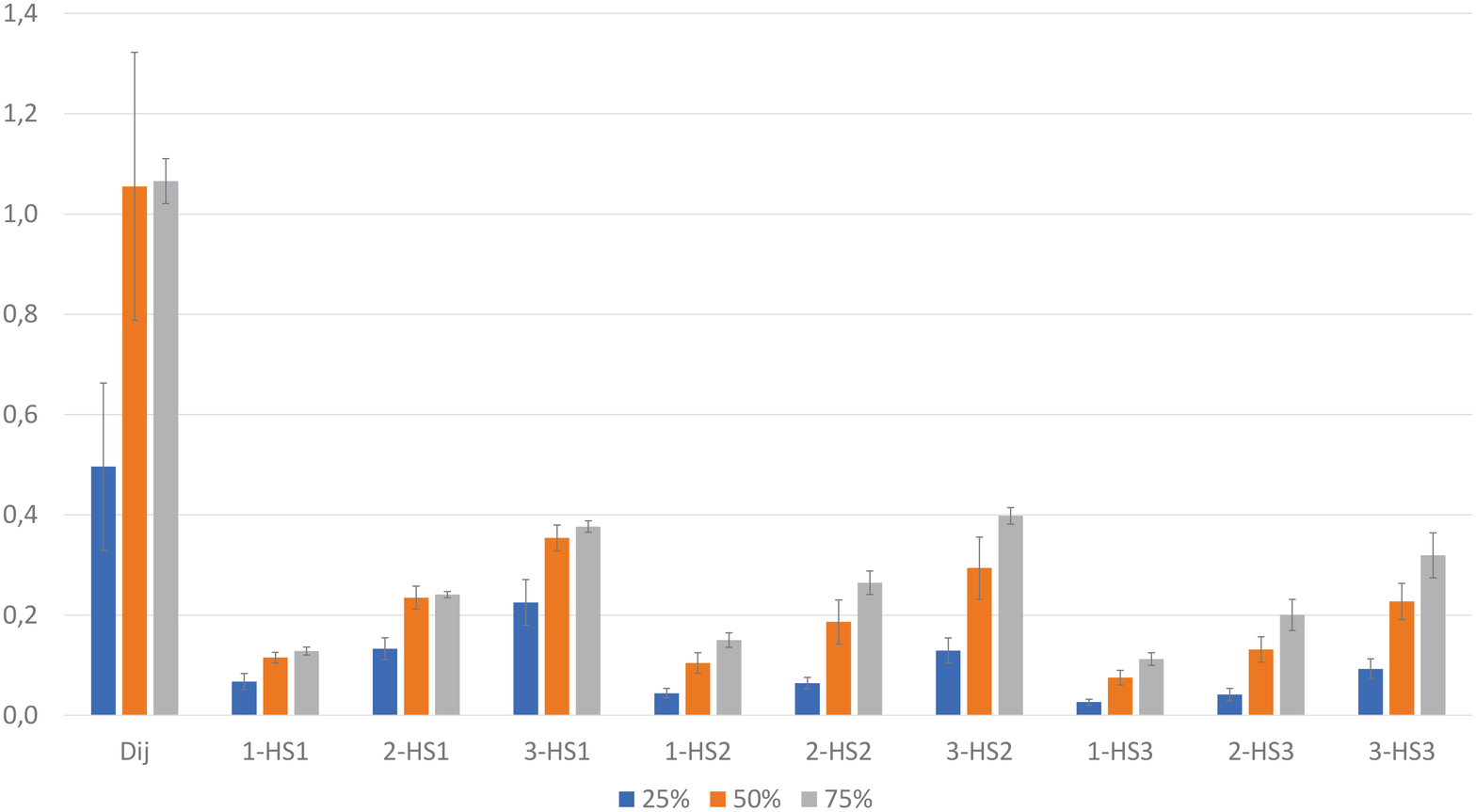}} \hfill
\subfloat[\label{fig-expres6}$n = 10000$, $r = 0.2$]{\includegraphics[width=0.45\textwidth]{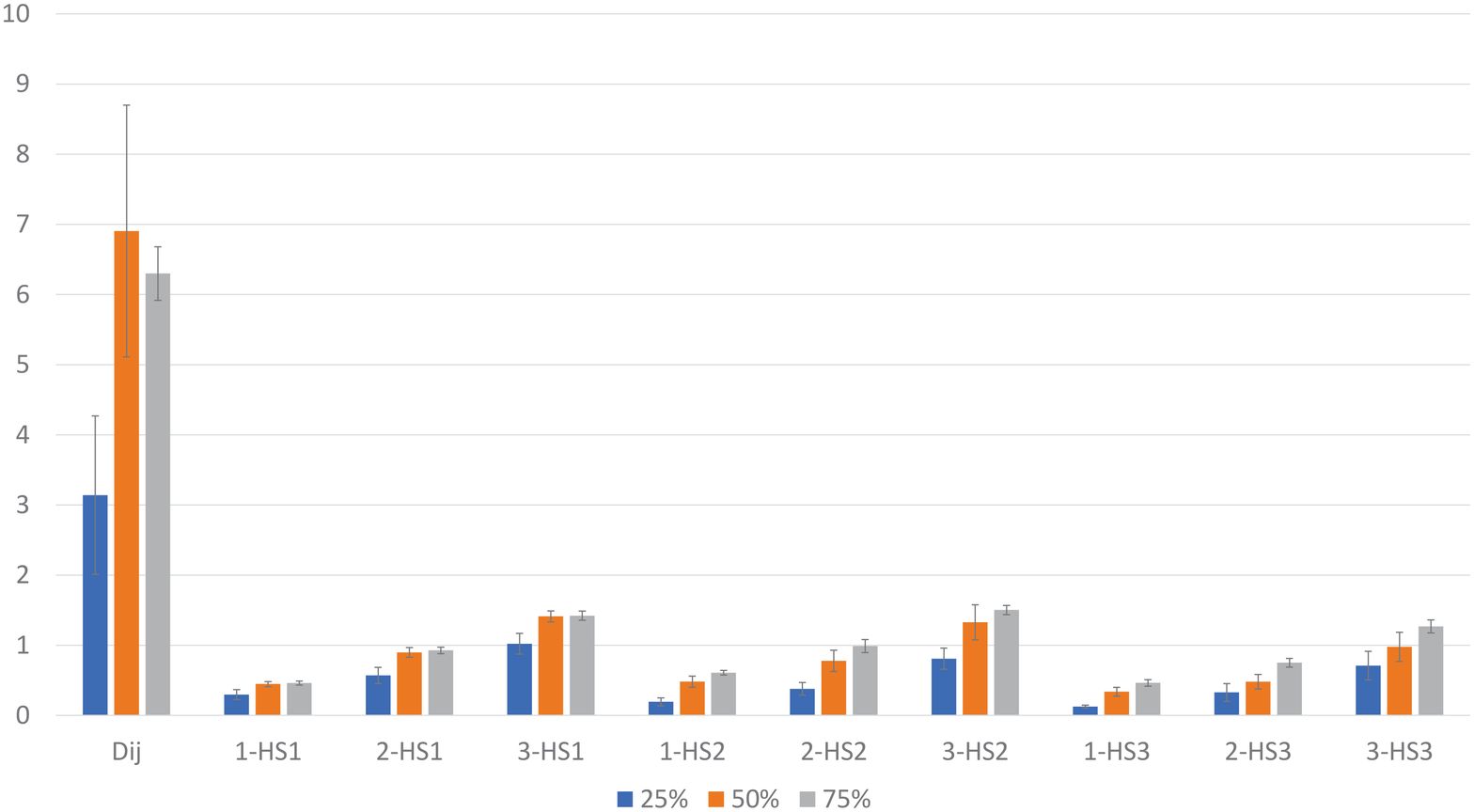}} \hfill
\subfloat[\label{fig-expres7}$n = 100000$, $r = 0.025$]{\includegraphics[width=0.45\textwidth]{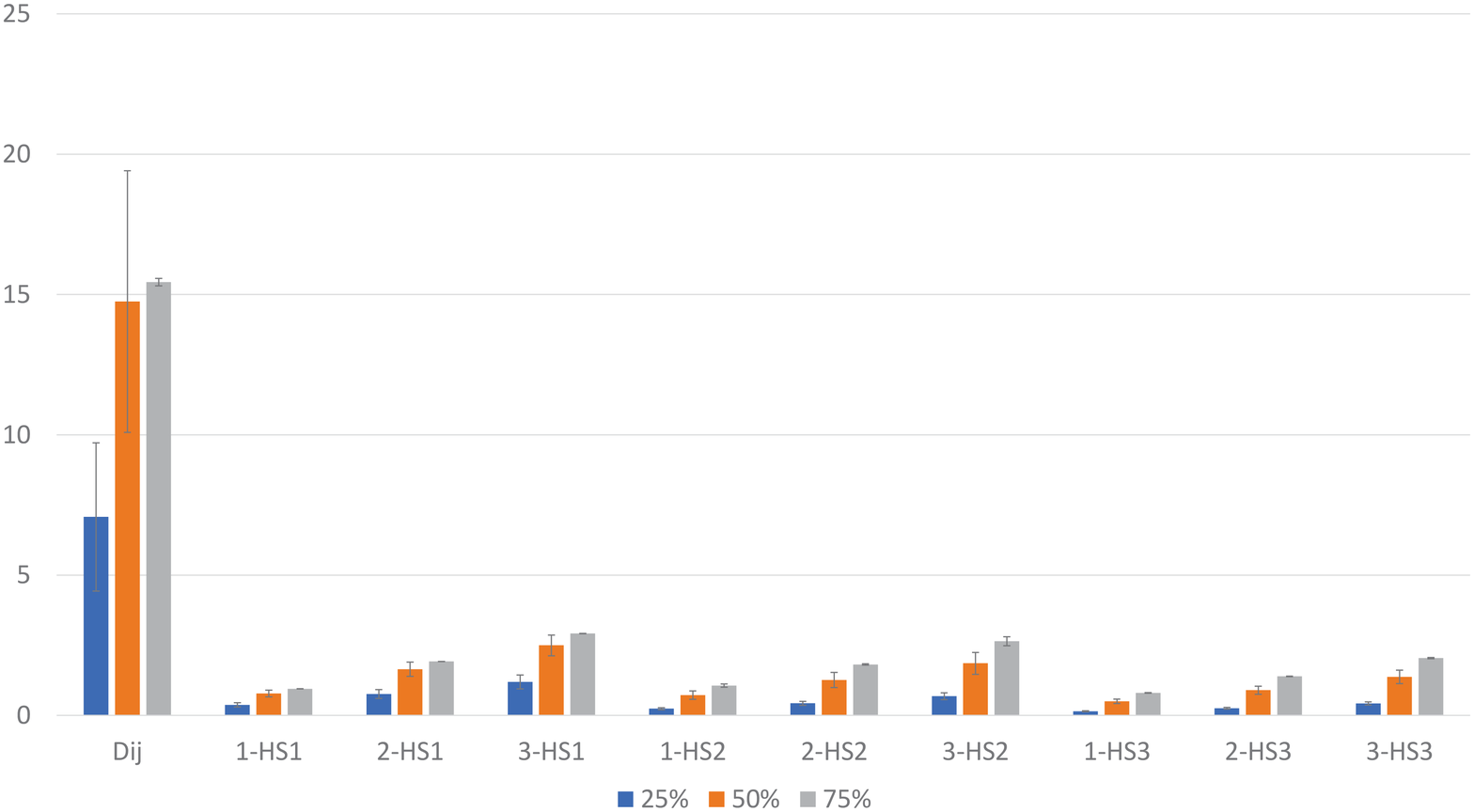}} \hfill
\caption{Time in seconds for the SP problem on the UDG. Average values and standard deviations.} \label{fig-expres57}
\end{figure}

\begin{figure}[h]
\centering
\subfloat[\label{fig-expres8}$n = 10000$, $r = 0.1$]{\includegraphics[width=0.45\textwidth]{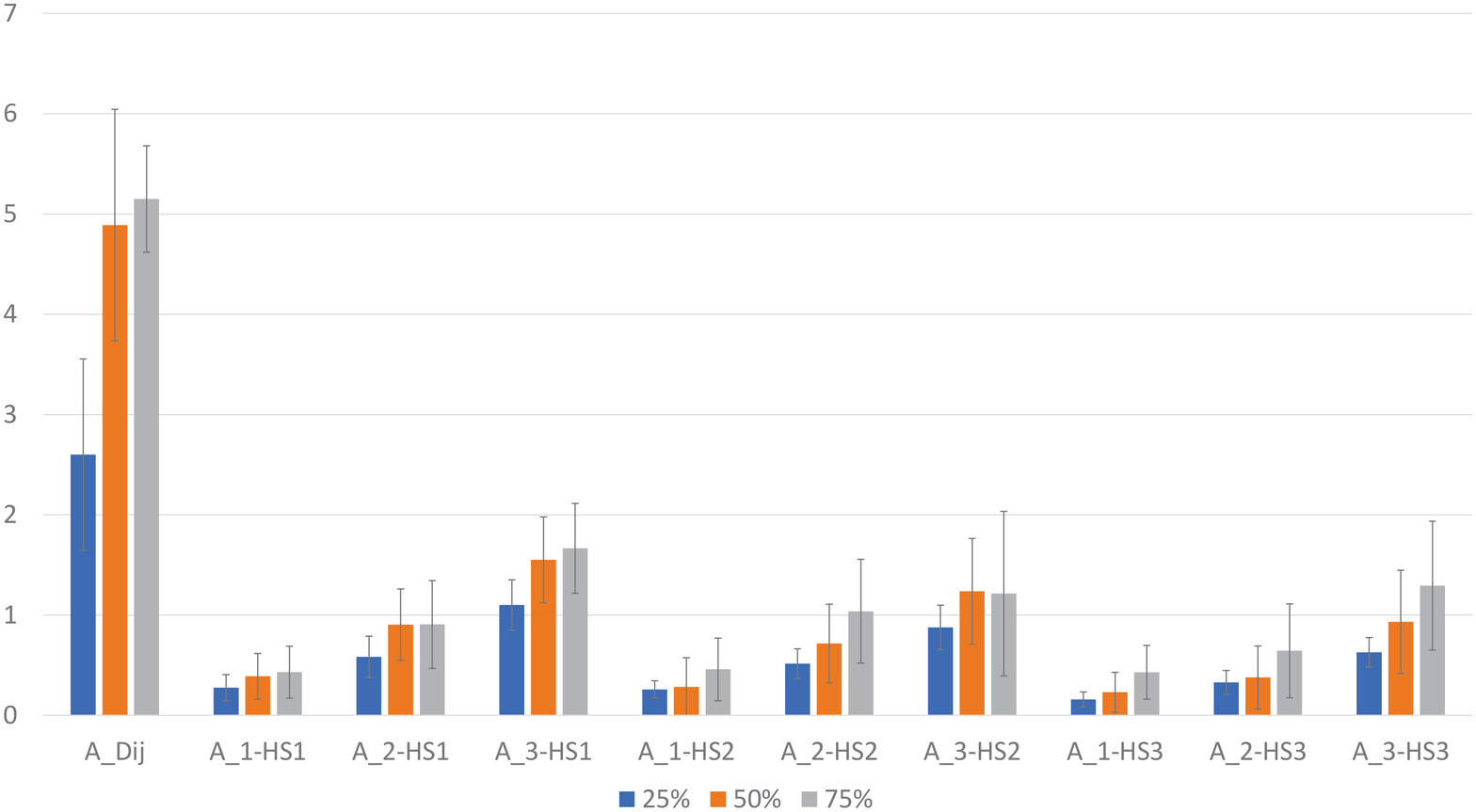}} \hfill
\subfloat[\label{fig-expres9}$n = 10000$, $r = 0.2$]{\includegraphics[width=0.45\textwidth]{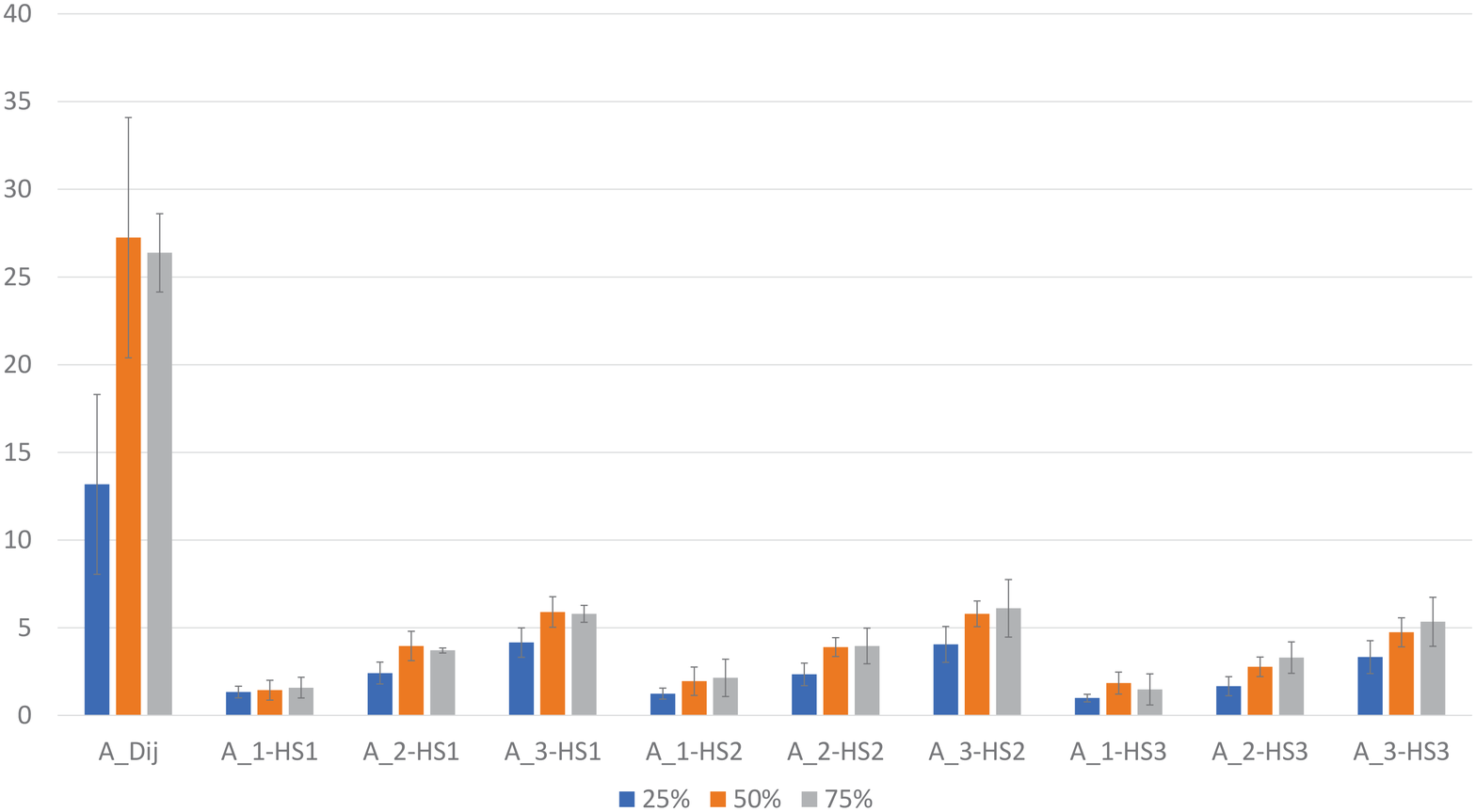}} \hfill
\subfloat[\label{fig-expres10}$n = 100000$, $r = 0.025$]{\includegraphics[width=0.45\textwidth]{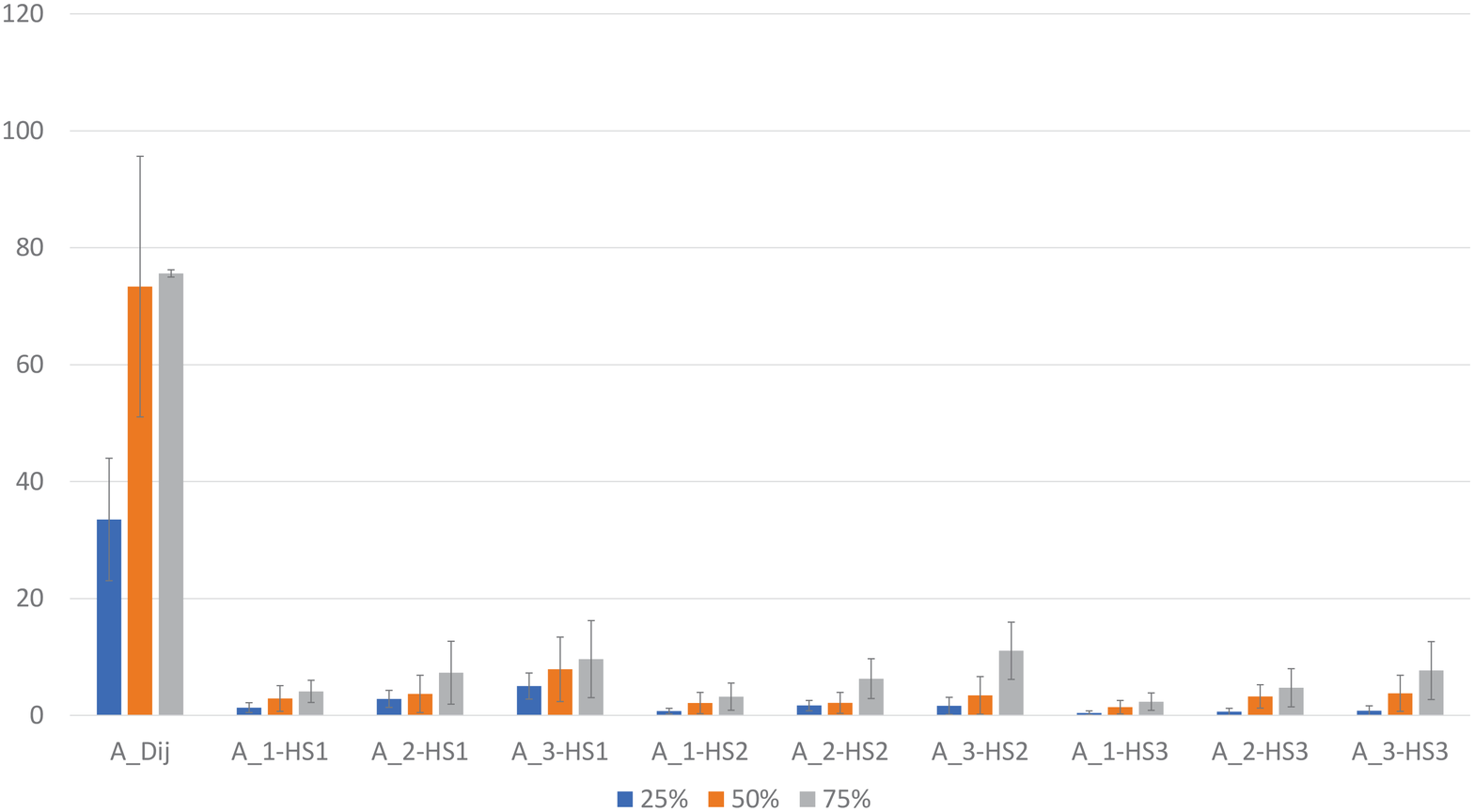}} \hfill
\caption{Time in seconds for the CSP problem on the UDG. Average values and standard deviations.} \label{fig-expres810}
\end{figure}

We also tested all the algorithms on the UDGs that were constructed as described above. Each time the points for the graph generation were randomly spread on the square with a side of the length 1. Graph density depends on two parameters: $n$ -- the number of vertices, and $r$ -- the \emph{disk} radius that defines connectivity between each pair of vertices. There were three UDG variants tested: (1) $n=10000$ and $r=0.1$,  (2) $n=10000$ and $r=0.2$, and  (3) $n=100000$ and $r=0.025$. As for SP so for CSP, all tested heuristics constructed almost optimal solution: the value of ratio of each algorithm never exceeded 1.002. Therefore, it is worth comparing only the running time. Fig. \ref{fig-expres57} presents the average running time for solving SP problem, and in Fig. \ref{fig-expres810} the average running time for solving CSP problem are shown. It can be noticed that for the both problems in the UDG usage of HS based heuristics instead of Dijkstra's algorithm is justified since they construct almost optimal solution an order of magnitude faster.

\section{Conclusion}
This paper considers NP-hard Constraint Shortest Path (CSP) problem, when to each arc of the given $n$-nodes graph two characteristics are assigned: cost and length, and it is required to find a min-cost length-bounded $s-t$ path between the given pair of nodes $s$ and $t$. The constraint on path length is accounted for by considering one edge weight equals to a linear combination of cost and length using a Lagrange multiplier as in \cite{Juttner:01}. By varying the multiplier value, a feasible solution delivers a minimum to the function with new weights. Then we are looking for a solution, not in the original graph but in the specially constructed hierarchical structures (HS). We show that the shortest path in the HS is constructed with $O(m)$-time complexity, where $m$ is the number of edges/arcs of the graph, and the approximate solution in the case of integer costs and lengths of the edges is found with $O(m\log n)$-time complexity. The a priori estimate of the algorithm's accuracy turned out to depend on the parameters of the problem and can be significant. Therefore, to evaluate the algorithm's effectiveness, we conducted a numerical experiment on the graphs of roads of megalopolis and randomly constructed unit-disk graphs (UDGs). The simulation shows that in the HS, a solution close to optimal one is built 10--100 times faster than in the methods using Dijkstra like algorithms to build a min-weight path in the original graph.


\begin{thebibliography}{18}

\bibitem{9dimacs} 9th DIMACS Implementation Challenge. http://www.dis.uniroma1.it/challenge9/ download.shtml

\bibitem{Ahuja:93} Ahuja R.K. et al.: Network Flows: Theory, Algorithms, and Applications. Prentice Hall, Inc., 1993

\bibitem{Cormen:00} Cormen T.H. et al.: Introduction to Algorithms. The MIT Press, Cambridge 2000

\bibitem{Garey:79} Garey M.S. and Johnson D.S.: Computers and Intractability: Guide to the Theory of NP-Completeness. (eds.) W.H. Freeman (eds.), New York, 1979

\bibitem{Goldberg:05} Goldberg A.V. and Chris H.: Computing the shortest path: A search meets graph theory. SODA '05 (2005).

\bibitem{Hassin:92} Hassin R.: Approximation schemes for the restricted shortest path problem. Mathematics of Operations Research \textbf{17}(1), 36--42 (1992)

\bibitem{Ishida:98} Ishida K. et al.: A delay-constrained least-cost path routing protocol and the synthesis method. In Proceedings of the 5th Int. Conf. on Real-Time Computing Systems and Applications. IEEE, 58--65 (1998)

\bibitem{Juttner:01} J\"{u}ttner A. et al.: Lagrange Relaxation Based Method for the QoS Routing Problem. IEEE INFOCOM 2001, 859--868 (2001)

\bibitem{Koster:14} Graphs and Algorithms in Communication Networks. Koster A.M.C., Mu\~{n}oz X. (eds.), Springer-Verlag Berlin Heidelberg, 2014

\bibitem{Handler:80} Handler, G. and I. Zang, I.:A dual algorithm for the constrained shortest path problem. Networks 10, 293--310 (1980)

\bibitem{Kuipers:02} Kuipers F.A. et al.: An overview of constraint-based path selection algorithms for QoS routing. IEEE Communications Magazine. \textbf{40}(12), 50 -- 55 (2002)

\bibitem{Lorenz:00} Lorenz D.H. et al.: Efficient QoS Partition and Routing of Unicast and Multicast. Proceedings of IWQoS 2000, 75--83 (2000)

\bibitem{Lozano:13} Lozano L.,and Medaglia A.L.: On an exact method for the constrained shortest path problem. Computers \& Operations Research. \textbf{40}, 378–-384 (2013)

\bibitem{Orda:99} Orda A.: Routing with End-to-End QoS Guarantees in Broadband Networks. IEEE/ACM Transactions on Networking. \textbf{7}(3), 365--374 (1999)

\bibitem{Pugliese:19} Pugliese L.D.P. et al.: The Resource Constrained Shortest Path Problem with uncertain data: A robust formulation and optimal solution approach. Computers and Operations Research. \textbf{107}, 140-–155 (2019)

\bibitem{Reeves:00} Reeves D.S. and Salama H.F.: A distributed algorithm for delay-constrained unicast routing. IEEE/ACM Transactions on Networking. \textbf{8}(2), 239--250 (2000)

\bibitem{Sun:98} Sun Q. and Langendorfer H.: A new distributed routing algorithm for supporting delay-sensitive applications. Computer Communications. \textbf{21}, 572--578 (1998)

\bibitem{Wang:14} Wang H. et al.: A Bio-Inspired Method for the Constrained Shortest Path Problem. The Scientific World Journal. V. 2014, Article ID 271280 (2014)

\bibitem{Wang:16} Wang S. et al.: Effective Indexing for Approximate Constrained Shortest Path Queries on Large Road Networks. Proceedings of the VLDB Endowment \textbf{10}(2), 61--72 (2016)

\bibitem{Wang:96} Wang Z. and Crowcroft J.: Quality-of-Service Routing for Supporting Multimedia Applications.
IEEE on Selected Areas in Communications. \textbf{14}(7), 1228--1234 (1996)

\bibitem{Widyono:94} Widyono R.: The design and evaluation of routing algorithms for real-time channels. Technical Report TR-94-024, University of California at Berkeley \& International Computer Science Institute (1994).

\bibitem{Xiao:05} Xiao Y. et al.: The Constrained Shortest Path Problem: Algorithmic Approaches and an Algebraic Study with Generalization. AKCE J. Graphs. Combin. \textbf{2}(2), 63--86 (2005)


\end{thebibliography}
\end{document}